\documentclass[runningheads]{llncs}
\usepackage[utf8]{inputenc}
\usepackage[T1]{fontenc}
\usepackage{graphicx} 
\graphicspath{ {./images/} }

\usepackage{newtxtext}

\usepackage[subtle]{savetrees}
\usepackage[english]{babel}
\usepackage{subcaption}
\usepackage{url}
\usepackage{colortbl}
\usepackage{amssymb}
\usepackage{stmaryrd}
\usepackage{amsmath}
\usepackage{mathrsfs}
\usepackage{amssymb}
\usepackage[left]{lineno}
\usepackage{xfrac}
\usepackage{nicefrac}
\usepackage[textsize=scriptsize,backgroundcolor=yellow!40,obeyFinal]{todonotes}
\usepackage[nonumberlist,acronym,sanitize=none]{glossaries}
\glsdisablehyper
\usepackage{comment}
\usepackage[pdftex, colorlinks=true, hyperfootnotes=true, hyperindex=true,
            plainpages=false, pagebackref=false, pdfpagelabels=true, pdfstartview=FitH,
            linkcolor=blue, citecolor=blue, urlcolor=blue,
            bookmarks, bookmarksopen, bookmarksdepth=3]{hyperref}
\usepackage[capitalise,nameinlink]{cleveref}
\crefname{algocf}{alg.}{algs.}
\Crefname{algocf}{Algorithm}{Algorithms}
\crefname{section}{Sect.}{Sects.}
\Crefname{section}{Section}{Sections}
\usepackage{tkz-base}
\usetikzlibrary{decorations.pathmorphing,trees,snakes,arrows,shapes,automata,petri}
\usepackage{paralist}
\usepackage{multicol}
\usepackage{booktabs}
\usepackage{enumitem}
\usepackage{multirow}
\usepackage{rotating}
\usepackage{etaremune}
\usepackage{marginnote}
\usepackage{mathtools}
\usepackage{mathabx}
\usepackage{adjustbox}
\usepackage{ifthen}
\usepackage[normalem]{ulem}
\usepackage{lineno}
\usepackage{soul}
\usepackage{floatrow}
\usepackage{listings}
\crefname{lstlisting}{Listing}{Listings}
\usepackage{lipsum}
\usepackage{makecell}
\usepackage{diagbox}
\usepackage[scientific-notation=false]{siunitx}
\usepackage{footmisc}
\usepackage{xspace}
\usepackage{ifdraft}
\usepackage{wrapfig}
\usepackage{orcidlink}%
\newacronym[\glslongpluralkey={Distributed Ledger Technologies}]{dlt}{DLT}{Distributed Ledger Technology}

\newacronym{ipfs}{IPFS}{InterPlanetary File System}

\newacronym{p2p}{P2P}{peer-to-peer}

\newacronym{abe}{ABE}{Attribute-Based Encryption}

\newacronym{maabe}{MA-ABE}{Multi-Authority Attribute-Based Encryption}

\newacronym{cpabe}{CP-ABE}{Ciphertext-Policy Attribute-Based Encryption}

\newacronym{ssl}{SSL}{Secure Sockets Layer}

\newacronym{dht}{DHT}{Distributed Hash Table}

\newacronym{mpc}{MPC}{Multi-party Computation}

\newglossaryentry{box}{ %
  name={Box},
  description={authority intialisation storage box},
  first={\glsentrydesc{box} (henceforth, \glsentrytext{Box} for short)},
  plural={Boxes},
  firstplural={\glsentrydesc{box}es (henceforth, \glsentryplural{box} for short)}
}

\newglossaryentry{rloc}{ %
	name={resource locator},
	description={a generic term to denote content-based links obtained via hashing (e.g., the IPFS link)},
}

\newacronym{cpmaabe}{CP-MA-ABE}{Ciphertext-Policy Multi-Authority Attribute-Based Encryption}

\newacronym{dk}{\textsl{dk}}{decryption key}
\newacronym{fdk}{\textsl{fdk}}{final decryption key}

\newacronym[\glslongpluralkey={Business Processes}]{bp}{BP}{Business Process}
\newacronym{bpi}{BPI}{Business Process Intelligence}
\newacronym{bpm}{BPM}{Business Process Management}
\newacronym{bpms}{BPMS}{Business Process Management System}
\newacronym{bpmn}{BPMN}{Business Process Model and Notation}
\newacronym{cpn}{CPN}{colored Petri net}
\newacronym{kpi}{KPI}{Key Performance Indicator}
\newacronym{ocbc}{OCBC}{Object-centric Behavioral Constraints}
\newacronym{soa}{SOA}{Service-Oriented Architecture}
\newacronym{pn}{PN}{Petri net}
\newacronym{wf}{WF}{workflow}
\newacronym{wfms}{WfMS}{Workflow Management System}
\newacronym{xes}{XES}{eXtensible Event Stream}
\newacronym{yawl}{YAWL}{Yet Another Workflow Language}
\newglossaryentry{task}{%
	name={task},description={the non-divisible, elementary activity}}

\newglossaryentry{promod}{%
	name={process model},description={the model of a process}
}
\def\LogAlph {\ensuremath{\Sigma}}
\newglossaryentry{logalph}{
	name={log alphabet},description={the process alphabet, as reflected in a log},%
	symbol={\LogAlph}}
\def\Evt {\ensuremath{e}}
\newglossaryentry{evt}{
	name={event},description={a record of an instantaneous fact during the process enactment},%
	symbol={\Evt}}
\def\Trc { \ensuremath{\tau} }

\newglossaryentry{trace}{
	name={trace},description={a sequence of \glsplural{evt}},%
	symbol={\Trc}}
\def\EvtLog {\ensuremath{L}}

\newglossaryentry{evtlog}{
	name={event log},description={a collection of \glstext{evttrace}s},%
	symbol={\EvtLog}}

\renewcommand\tabcolsep{5pt}
\newcolumntype{d}{>{\columncolor{gray!10}}c}
\newcolumntype{m}{>{\columncolor{gray!10}}l}
\newfloatcommand{capbtabbox}{table}[][\FBwidth]
\newenvironment{iiilist}%
{\begin{inparaenum}[\itshape(i)\upshape]}%
{\end{inparaenum}}

\RequirePackage{xparse}
\NewDocumentEnvironment{AuthNote}{+o+o}{%
	\IfValueT{#2}{\marginnote{\scriptsize{#2}}}%
	\begin{scriptsize}
		\colorbox{gray}%
		{\color{white} Note\IfValueT{#1}{ (#1)}:}%
		\quad%
		\color{brown}
}{%
	\normalcolor
	\end{scriptsize}
}

\RequirePackage{pifont}

\RequirePackage{lipsum}
\newcommand{\LipsumGray}[1][]{{\color{gray}\ifthenelse{\equal{#1}{}}{\lipsum}{\lipsum[#1]}}}

\RequirePackage{siunitx}
\newcolumntype{D}[1]{S[
	table-omit-exponent,
	round-mode=places,
	round-integer-to-decimal,
	round-precision={#1}]} 

\begin{document}

\title{Transforming Credit Guarantee Schemes with Distributed Ledger Technology}

\author{ 
	Sabrina~Leo\inst{1}\orcidlink{0000-0002-5157-2135} 
	\and
	Andrea~Delle~Foglie\inst{1}\orcidlink{0000-0002-8946-6338}
        \and
	Luca~Barbaro\inst{1}\orcidlink{0000-0002-2975-5330}
        \and
	Edoardo~Marangone\inst{1}\orcidlink{0000-0002-0565-9168}
	\and
	Ida~Claudia~Panetta\inst{1}\orcidlink{0000-0001-7019-3993} 
	\and
	Claudio~Di~Ciccio\inst{2}\orcidlink{0000-0001-5570-0475}
}
\authorrunning{ Leo et al.}

\institute{ 
	Sapienza University of Rome, Rome, Italy,
	\email{\href{name.surname@uniroma1.it}{name.surname@uniroma1.it}}   
	\and
	Utrecht University, Utrecht, Netherlands,
	\email{\href{mailto:c.diciccio@uu.nl}{c.diciccio@uu.nl}}
}
\maketitle

\begin{abstract}
 Credit Guarantee Schemes (CGSs) are crucial in mitigating SMEs' financial constraints. However, they are renownedly affected by critical shortcomings, such as a lack of financial sustainability and operational efficiency. Distributed Ledger Technologies (DLTs) have shown significant revolutionary influence in several sectors, including finance and banking, thanks to the full operational traceability they bring alongside verifiable computation. Nevertheless, the potential synergy between DLTs and CGSs has not been thoroughly investigated yet. This paper proposes a comprehensive framework to utilise DLTs, particularly blockchain technologies, in CGS processes to improve operational efficiency and effectiveness. To this end, we compare key architectural characteristics considering access level, governance structure, and consensus method, to examine their fit with CGS processes. We believe this study can guide policymakers and stakeholders, thereby stimulating further innovation in this promising field.
\end{abstract}

\section{Introduction}
\label{introduction}
In the credit industry, a guarantee reduces the financial risk of loans by providing a lending bank with additional assurances, increasing the likelihood of capital.%
\footnote{European Central Bank: ``What is collateral?'' (2016): \href{https://www.ecb.europa.eu/ecb/educational/explainers/tell-me/html/collateral.en.html}{\nolinkurl{www.ecb.europa.eu/ecb/educational/ explainers/tell-me/html/collateral.en.html}}
}
It also signals the borrower’s intention to repay, reducing monitoring costs. Credit Guarantee Schemes (CGSs) act as a third-party provider of credit risk mitigation to lenders. By effectively reducing the risk associated with lending, they incentivise financial institutions to extend credit to Small and Medium-sized Enterprises (SMEs), thereby enhancing their financial stability and prospects for growth.
CGSs received increased attention during the COVID-19 pandemic as government credit guarantee programmes improved banks' balance sheets and facilitated credit extensions to creditworthy businesses. The positive impact on post-crisis productivity was attributed to expanded coverage and simplified procedures in these schemes~\cite{Post-COVID-19,From-hibernation-to-reallocation}.

However, the literature has highlighted the financial sustainability issues of CGSs, particularly the risk of high default rates, which can force them to pay significant rewards~\cite{Ayyagari}. A common criticism concerns the lack of transparency between the parties involved in the transaction (from customer to CGS and from bank to CGS), which leads to problems such as moral hazard for borrowers or adverse lender selection. Furthermore, the correct assessment of the customer's creditworthiness and the procedure's cost-effectiveness influence the credit guarantees' additional attributes. Therefore, CGS's potential also depends on the sustainability and efficiency of its operational processes.

Emerging Distributed Ledger Technologies (DLTs), such as blockchain, open up a chance for processes involving third parties to be trustworthily executed~\cite{Zetzsche}, particularly in scenarios where there is a lack of mutual trust and confidence between the parties.
%
Specifically, by leveraging blockchain's main features, including operational traceability and transparency, it is possible to minimise the risk of fraudulent activities, enhance customer trust, lower operational costs through process automation support, and reduce downtime.
%

In this paper, we describe the main challenges and opportunities of blockchain technology adoption for addressing the inherent weaknesses of traditional CGSs. We also suggest a complete framework that includes all the important steps in the CGS operational process, from the application to possibly enforcing safety measures. This framework will help identify key areas that could benefit from integration with DLTs. Given the specificity of the procedures, the actors involved, and data sensitivity in transactions, our paper raises questions about the appropriate blockchain architecture that can improve transparency, security, and automation throughout the entire CGS lifecycle.
%
Our paper provides a roadmap for policymakers, financial institutions, and stakeholders who seek to modernise and optimise CGS mechanisms. 
To the best of our knowledge, this is the first endeavour exploring a synergy between DLT and CGS. We hope it can serve to stimulate further research and innovation in this nascent yet promising field.

In the following, \cref{background}, outlines the core ideas that form the basis of our proposal. In \cref{discussion}, we discuss why given blockchain configurations are suitable our setting. \Cref{scenario} describes their possible application in the credit guarantee life cycle. Finally, \cref{conclusion} concludes the paper.
\section{Background}
\label{background}
%
\subsection{Distributed Ledger Technologies}
\textbf{Distributed Ledger Technologies (DLTs)} comprehend protocols designed to enable the transactions' storage, processing, and validation within a decentralised network of interconnected peers. A distributed ledger is a registry replicated over a network of computing systems (henceforth, \emph{nodes}) that records the sequence of transactions containing the transfer of value or data between sender and recipient accounts~\cite{DiCiccio/DLTBanking2023:BlockchainDistributedLedger}. Its distributed architecture obviates the necessity for centralised authorities or intermediaries in data management. Every transaction within this framework features a timestamp and a unique cryptographic signature to prove that the account's owner issued it. Signatures are generated utilising a public/private key scheme, where each user possesses an account associated with a distinct address to which these keys are linked. This setting guarantees pseudonymity. We recall that pseudonymity differs from anonymity in that a unique identifier (in this case, the account number) is associated with the same individual, singled out across the registered data. The list of the shared transactions contained in the ledger is accessible to all network participants (owing to its public nature) so that everyone can verify that the digital signature belongs to the owner and trace all the transfers from or to an account. 

\textbf{Blockchain}, a widely known class of DLT, entails aggregating of ledger segments into blocks, which are then sequentially linked backwards to form an immutable chain. The integrity of DLTs, including blockchain, is fortified through a confluence of cryptographic methodologies and the distributed validation of transactions. In blockchains, every block is linked to the preceding one by keeping in the former a fingerprint number identifying the latter. This fingerprint is yielded by a mathematical function called \emph{hashing}. Hashing produces a number (the hash). It is deterministic (given the same data, the same hash is returned) but one-way (hashes cannot be reverse-engineered to retrieve the input). Altering a bit turns the associated hash into a completely different number. Well-known blockchain platforms such as Bitcoin~\cite{Nakamoto/2008:Bitcoin:APeer-to-PeerElectronicCashSystem}, Ethereum~\cite{Wood/2018:Ethereum}, and Algorand~\cite{Chen.Micali/TCS2019:Algorand} necessitate transaction fees to allow for the submission and processing of transactions. Fees serve as an economic incentive to remunerate the computing systems in the network which include and maintain the infrastructure up and running.
To ensure a single, reliable and generally acknowledged version of the transaction record, blockchains rely on consensus algorithms. These algorithms guarantee that all participants agree on the content of the ledger and that only trusted nodes can add new information. Proof of Work (PoW)~\cite{PoW} and Proof of Stake (PoS)~\cite{PoS} are two renowned consensus mechanisms used to validate transactions and add new blocks to a blockchain. 
%
Blockchain platforms such as Ethereum and Algorand offer the capability to execute \textbf{smart contracts}, namely executable programs that are deployed, stored, and run within the blockchain environment~\cite{Dannen/2017:IntroducingEthereumandSolidity}.
Smart contracts' operations are triggered by transactions, hence without the involvement of a trusted third party. Therefore, the history of invocations can be verified, replayed and trusted throughout the whole network. As with transactions, the execution of smart contract code is subject to costs that fall under the name of \emph{gas} in the Ethereum nomenclature. These costs depend on the complexity of the invoked code and the amount of data exchanged and stored. To reduce the invocation costs of smart contracts, external peer-to-peer systems are typically employed to save larger chunks of data. \textbf{Inter Planetary File System (IPFS)}\footnote{\href{https://ipfs.tech/}{\nolinkurl{ipfs.tech}}} is one of the most commonly used distributed systems to store data across multiple peers without the involvement of any central authority or trusted organisation. The address of an IPFS file is based on the hash of the file itself, so that the contents and its locator are immutably bound. The locator is thus typically saved by smart contracts, which permanently store the link to that resource on-chain.

Blockchains are often classified according to transactability (public or private) and consensus mechanism (permissionless or permissioned).
If a blockchain is \textbf{public}, every node is endowed with the capability to initiate and monitor transactions. This fosters a transparent environment readily accessible to any users. In contrast, the blockchain is \textbf{private} if only selected nodes are entitled to participate in transaction initiation and observation. If any node in the network can participate in the consensus decision-making process, the blockchain platform is \textbf{permissionless}. Instead, if only specific nodes can determine the next status of the blockchain, it is \textbf{permissioned}.
As a result, blockchain platforms are typically classified into four different categories: public permissionless, public permissioned, private permissionless and private permissioned.
These characteristics discriminate the suitability of a blockchain platform to a given application scenario.

\subsection{Credit Guarantee Schemes}
SME credit markets are known for experiencing financial constraints and imperfections such as information asymmetries, insufficient or absence of recognised collateral, high transaction costs for small-scale lending, and perceived high risk, all contributing to suboptimal credit allocation. In this scenario, CGSs are crucial in offering lenders third-party credit risk mitigation to enhance SMEs' credit availability, acting as a strategic component in the market for credit distribution, and minimising distortions in credit markets. Moreover, CGS can operate as a tool at the disposal of policymakers to help SMEs finance during an economic downturn when risk aversion rises and a credit crunch is expected~\cite{WorldBank}. 

Numerous stakeholders partake and engage in the proceedings. 
\textbf{Government agencies} or Ministries oversee public CGSs by providing policy guidance, funding, and regulatory frameworks. From their standpoint, CGSs are an integral part of national development strategies, promoting the growth of SMEs and financial inclusion~\cite{Schich}. 
\textbf{Credit Guarantee Institutions (CGIs)} manage CGSs, evaluate credit risk, provide guarantees, and handle claims. They work with lenders to reduce credit risk and promote financing for SMEs~\cite{Leone2013}. 
\textbf{Credit institutions} such as banks and microfinance institutions work with CGSs and CGIs to provide lower-risk loans to SMEs. The ultimate beneficiaries of CGS-guaranteed credit are \textbf{SMEs} (generally speaking, the borrowers), who use it to finance their operations and promote investment, growth, and job creation. CGS enhances credit rating to facilitate business expansion. \textbf{Investors and shareholders}, according to the type of funds provided (i.e., Mutual Guarantee Institutions~\cite{delafuente}, Corporate Guarantee Schemes, Public Guarantee Schemes), expect different types of returns when supporting the development of SMEs. \textbf{External auditors} independently evaluate the financial status and adherence to CGI regulations by assessing their efficiency in supporting the funding of SMEs.
One of the multiple challenges associated with evaluating a CGS concerns providing useful information for the decision-making of agents who maintain an interest in the entity~\cite{Haskell2017}. Not all stakeholders have a similar degree of information at their disposal. On the one hand, internal stakeholders, such as managers, have unlimited access to information. On the other hand, external stakeholders, such as resource contributors and others, need to receive quality information. An important group of stakeholders includes banks, as they mobilise the guaranteed resources towards the final beneficiary. 
They require information that allows them to analyse the financial sustainability of the CGS and its ability to meet its commitments and comply with Basel III requirements if applicable.\footnote{European Central Bank: ``Sound practices in Counterparty Credit Risk governance and management'' (2023). \href{https://www.bankingsupervision.europa.eu/legalframework/publiccons/pdf/ccr_report/ssm.ccrgovernancemanagement_202306.en.pdf}{\nolinkurl{www.bankingsupervision.europa.eu/legalframework/publiccons/pdf/ccr\_report/ ssm.ccrgovernancemanagement\_202306.en.pdf}}%
} In addition, even more relevantly, throughout the life cycle of the guarantee, banks and CGIs need to constantly share reliable and up-to-date information on the condition of the borrower they are helping to finance. 
In this crowded stage, the presence of a mechanism that standardises data and optimises information sharing may protect the interests of all stakeholders.

\section{Selecting a Blockchain Configuration for CGS Support}
\label{discussion}
This section discusses advantages and challenges that the adoption of blockchain platforms can bring within the CGS scenario.

Information stored on a ledger is tamper-proof, persistent, and non-repudiable.
The layer of trust that blockchain infrastructures thereby provide is a key feature in potentially semi-trusted, competitive or untrusted multi-party business scenarios~\cite{DBLP:conf/bpm/WeberXRGPM16}. Therefore, they are apt for the case of a complex scenario such as that of CGS processes shown in \cref{background}. 
Furthermore, the need to encode the cooperative behaviour of the system supporting CGSs call for the use of environments that support smart contracts. Existing mechanisms for the semi-automation of collaborative processes based on smart-contract enabled platforms provide a fertile ground for our specific use case~\cite{DiCiccio.etal/InfSpektrum2019:BlockchainSupportforCollaborativeBusinessProcesses,DBLP:journals/fgcs/CorradiniM0P0023}. 
As outlined in \cref{background}, finding a suitable platform category is a fundamental strategic decision to design a system based on blockchain~\cite{Wuest.Gervais/CVCBT2018:DoyouNeedaBlockchain}. In the following, we outline some key criteria guiding this choice in our context. 

A \textbf{public permissionless} blockchain scheme like that of Ethereum and Algorand puts the secrecy of confidential data at risk and cannot be used in this scenario except with the necessary encryption of those reserved pieces of information. A common approach is to encrypt them via symmetric keys individually provided for readers~\cite{NIST/FIPS197-2001:AES,symmetric}.
However, this solution entails the additional overhead of managing keys at an individual level (which undermines scalability), keeping track of the shared ones (at the risk of breaching confidentiality), or having multiple duplicates of the same message with different encryptions (thus creating potential security issues caused by inconsistent copies).
A promising alternative in that scenario is to decouple the attribution of keys from the individuals retaining them as proposed in the literature with frameworks such as Control Access via Key Encryption (CAKE)~\cite{Marangone.etal/BPM2022:CAKE} and Multi-Authority Approach to Transaction Systems for Interoperating Applications MARTSIA~\cite{Marangone.etal/EDOC2023:MARTSIA}. These solutions, grounded in Attribute-Based Encryption (ABE)~\cite{CP-ABE}, resort to IPFS to store sensitive information and policies that determine data access based on the certified roles of the actors (e.g., whether an account is linked to a customer or not, a bank or not, and so forth). Data are thus encrypted and decrypted with keys that decouple the individuals from their access grants. 

\textbf{Public permissioned} schemes (see, e.g., EOSIO%
\footnote{%
\href{https://github.com/EOSIO/Documentation/blob/master/TechnicalWhitePaper.md/}{\url{github.com/EOSIO/Documentation/blob/master/TechnicalWhitePaper.md}}})
have been developed to implement more restrictive consensus mechanisms on public networks. 
Public permissioned schemes suffer from the same issue as the public permissionless ones in terms of confidentiality constraints as the visibility of the transaction contents remains public. In addition, it may call for additional decision processes to decide who manages the ledger and leads consensus. Alternatively, organisations may opt for the implementation of private blockchains.

\textbf{Private permissioned} networks such as Hyperledger Fabric~\cite{Androulaki.etal/ERCIMn2017:PermissionedBlockchainsandHyperledgerFabric} exert control over the nodes with mechanisms including the verification of their actual identities, in line with the Know Your Customer (KYC) principles~\cite{DBLP:journals/bise/MoyanoR17}.
Platforms of this sort often include mechanisms for Role Based Access Control~\cite{Baset.etal/2018:HyperledgerFabric}.
Another advantage of private blockchain networks is that the transaction costs, if any, are not bound to conversion into fiat currencies, as in the case of Ether, Algos, or Bitcoin, which is beneficial for limiting infrastructural costs.
Only a subset of nodes participate in the consensus, so the content of the ledger is under the control of authoritative entities, too.
These aspects are beneficial to preserve data confidentiality and system security. However, they demand an initial contracting phase and potentially conflicting subsequent management of who is in control of the ledger due to the permissioned nature of the infrastructure. In a setting such as that of CGSs, wherein a multitude of actors with cooperative scenarios and competing interests are involved, this situation might turn out to be undesirable.

\textbf{Private permissionless} blockchains, like LTO Network%
\footnote{%
\href{https://ltonetwork.com/}{\url{ltonetwork.com}}}
have recently gained momentum. 
Similarly to the public permissionless setting, any node can take part in the consensus algorithm.
However, unlike a public blockchain, only selected nodes are allowed to issue and read transactions. This hybrid nature is promising in our scenario as the distributed responsibility over the publication of transactions entails that no complex management procedures are in place to determine the entities in control of the ledger's content.
At the same time, only nodes associated with certified actors enter the network. In our scenario, this implies that the exchanged information is accessible to the sole actors involved in the CGI.

\section{Integrating Blockchain into the CGS Workflow}
\label{scenario}
\begin{figure*}[tb]
	\includegraphics[width=0.6\textheight]{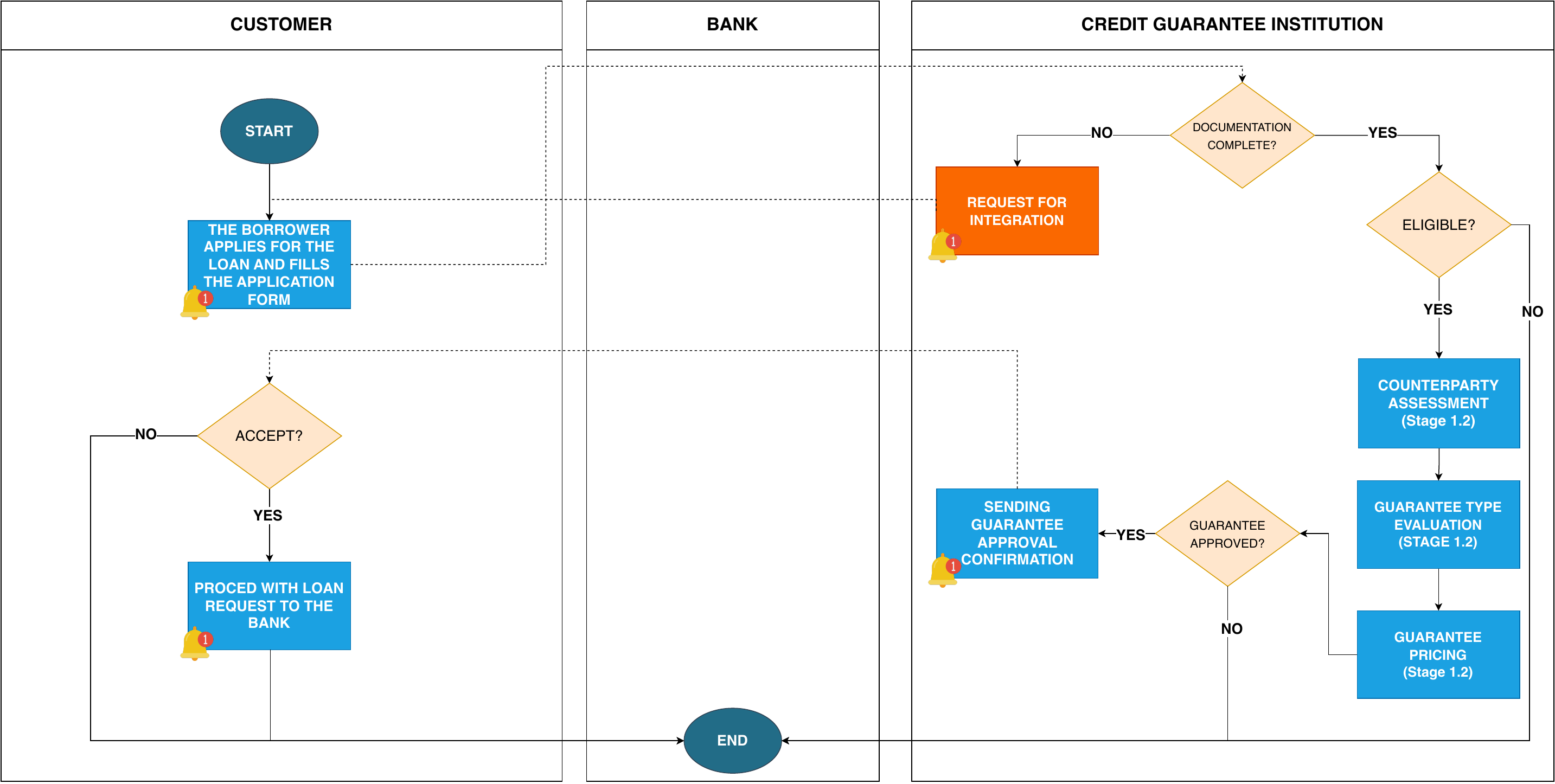}
	\caption{Individual Guarantee Scheme - Ex-ante application flowchart}
	\label{fig:ex-ante}
\end{figure*}

\begin{figure*}[tb]
	\includegraphics[width=0.6\textheight]{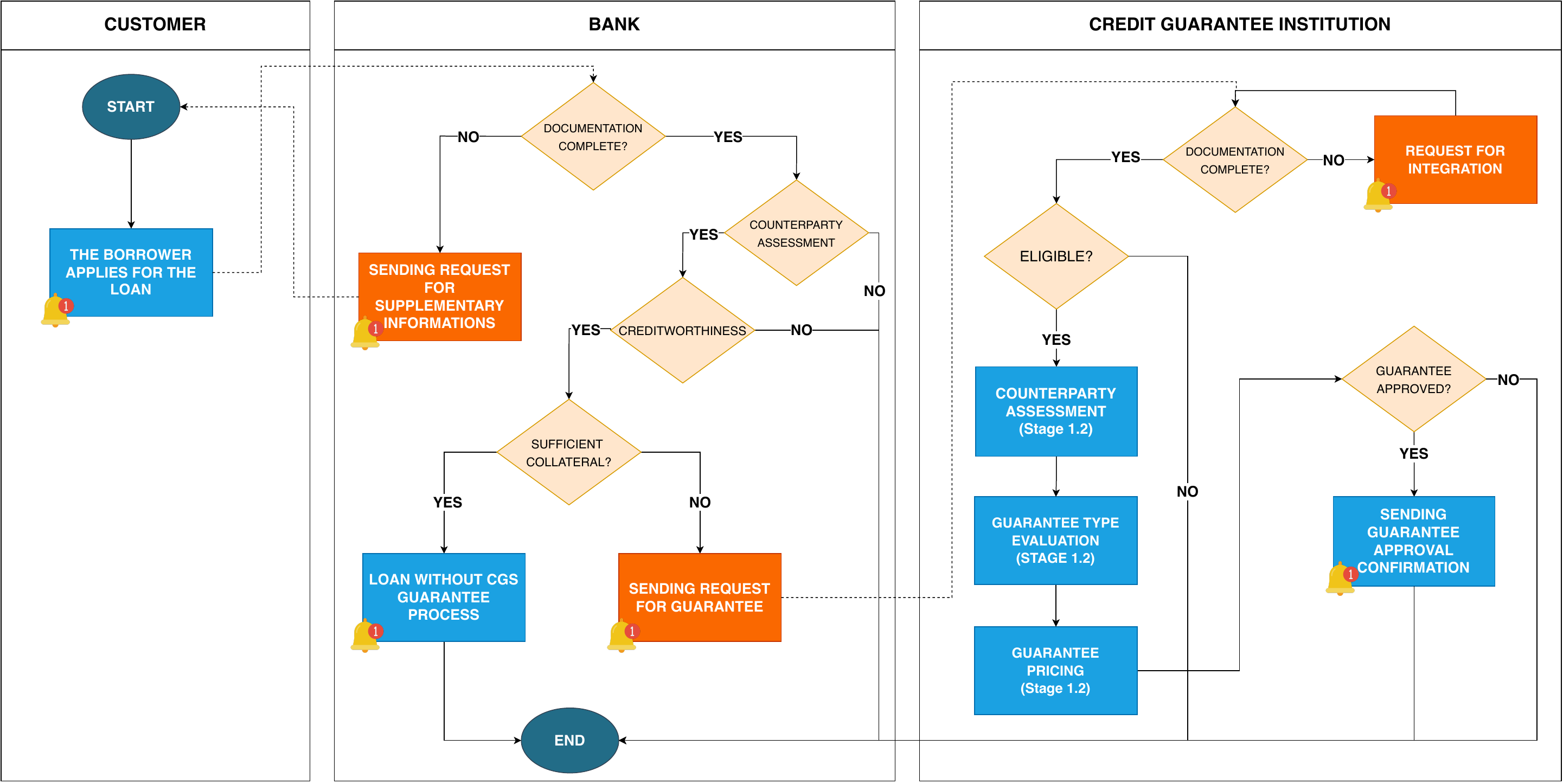}
	\caption{Individual Guarantee Scheme - Ex-post application flowchart}
	\label{fig:ex-post}
\end{figure*}

\begin{figure*}[tb]
	\includegraphics[width=0.6\textheight]{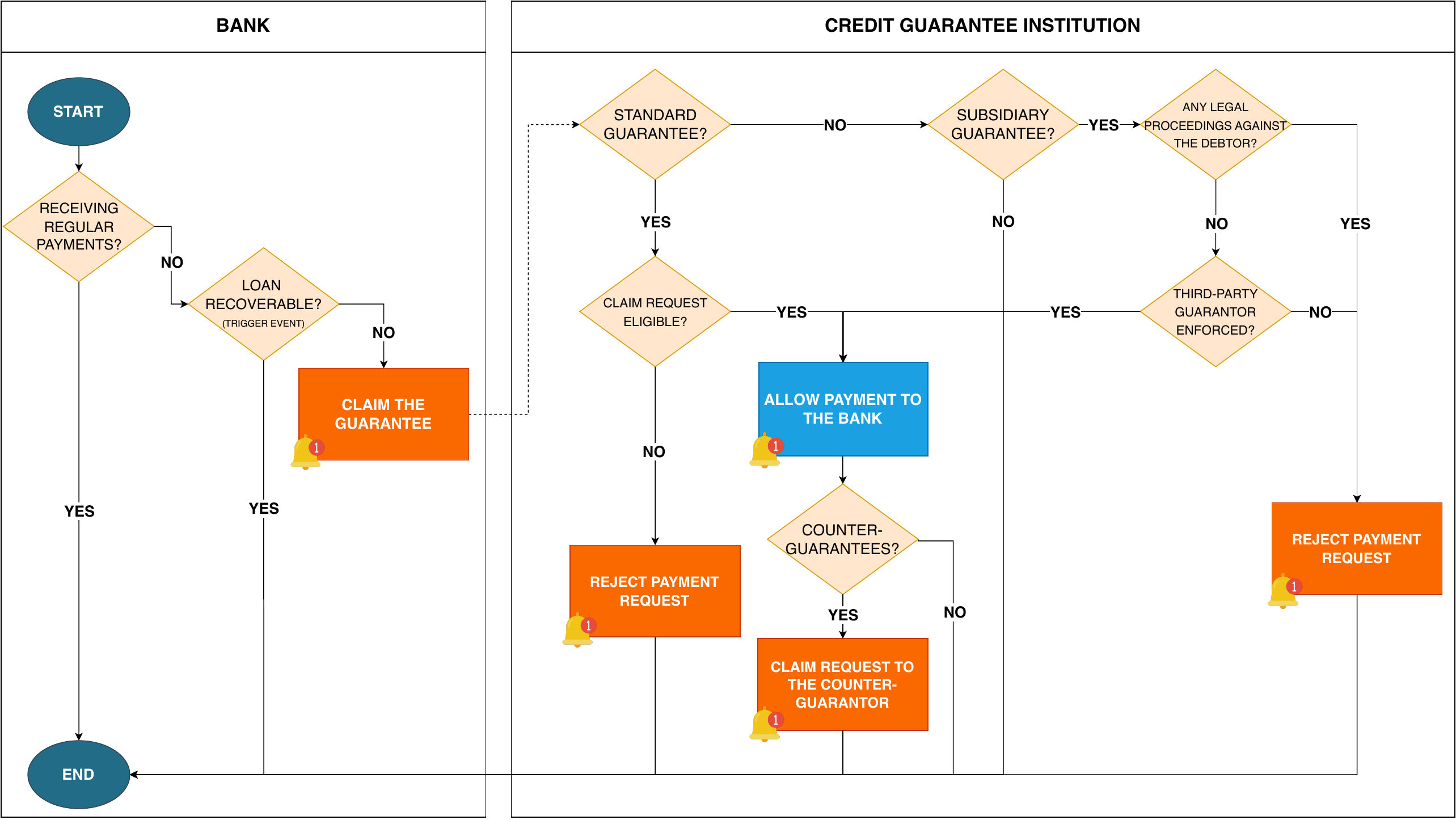}
	\caption{Credit Guarantee Scheme - Enforcement}
	\label{fig:enforcement}
\end{figure*}

In this section, we represent the key stages of the credit guarantee life cycle, which includes three actors: 
\begin{iiilist}
\item the borrower (the debtor of the loans or SMEs),
\item the bank (the lender)
and 
\item the Credit Guarantee Institution (CGI, the guarantor)~\cite{deelen2004guarantee}.
\end{iiilist}%
%
%
In particular, we examine individual guarantee schemes, focusing on two specific aspects related to the timing of guarantee requests in relation to banks, namely whether these are requested ex-ante or ex-post. In these schemes, the CGI evaluates each loan application a lender submits to the bank either before (ex-ante) or after (ex-post).
%
In \Cref{fig:ex-ante,fig:ex-post,fig:enforcement},  we use flowcharts to highlight the key steps of the processes where the implementation of blockchain-based solutions can help improve efficiency and ensure process transparency. 
In addition, bell-shaped icons highlight the points where smart contracts may step in. 

\Cref{fig:ex-ante} illustrates an example of an ex-ante application process. The smart contract and typical blockchain features can be adopted to seamlessly align with the \textbf{Know Your Customer (KYC)}~\cite{DBLP:journals/bise/MoyanoR17} process required by the CGI, starting with the receipt of the application form. Smart contracts can automate, verify, and exchange KYC data safely and transparently by utilising certain configurations of blockchain-based distributed ledgers, allowing customers to submit their information.

Once stored on the blockchain, data cannot be altered or deleted, ensuring protection against fraud and unauthorised access. Furthermore, the network's interoperability can enable financial data sharing with the bank after the guarantee is accepted, enabling the loan application to be submitted. 
The smart contract can automate the sharing of KYC data between financial institutions, minimising the workload on both parties and maintaining compliance with cross-border requirements. Moreover, it can significantly speed up the KYC process, leading to quicker onboarding of new clients and lowering administrative costs, enhancing CGI's competitiveness. The contract code can automatically verify the information borrowers submit to approve the KYC status immediately without manual verification and indicate the absence of required additional data if need be. 
The CGI can efficiently interact with the borrower using a notification system based on the blockchain network. Upon compliance with the predefined rules and regulations of the smart contract (i.e., the successful outcome of the procedure), the automatic loan request can be submitted to the bank. This proposal focuses on applying blockchains to the CGI collateral issuance process. When considering bank valuation, interoperability must be deemed to allow all parties to participate in the blockchain network and possibly incorporate the loan provisioning procedure.
%

In the ex-post application process (\cref{fig:ex-post}), the bank first evaluates the loan application and the presence of sufficient collateral. The bank can expedite the inquiry process by sending the request directly to the CGI if necessary, as the borrower has already been checked and deemed eligible for the loan. In contrast with the ex-ante scenario, all subjects can be involved from the beginning. In this context, the effectiveness of blockchains and smart contracts depends on their ability to automate and optimise processes by exploiting the technology's unique characteristics and on the interoperability that would allow cooperation and data sharing between the different entities involved. Additionally, the immutability of DLTs improves openness and auditability among participants by recording KYC data on the blockchain, making it readily available to authorised parties. As with the ex-ante process, setting rules and regulations in the smart contract on the blockchain network can speed up the onboarding process by allowing the CGI to establish criteria in advance and verify them automatically.
After the guarantee is approved and the risk-sharing agreement between the CGI and the bank has been achieved, the last condition to proceed with the loan disbursement is the guarantee fee payment verification. Once this condition has been respected, issuing the guarantee certificate is possible. As mentioned above in the flowcharts, in this stage, the platform's interoperability is essential since all the actors are involved in the process, and with advanced applications, it should be possible to integrate a smart loan contract. The smart contract previously deployed in \cref{fig:ex-ante} and \cref{fig:ex-post} can also include this part of the process by assuming that the CGI approves the guarantee. As previously mentioned, thanks to the smart contract preset rules and regulations, most of the process is completely automated, even if negotiating the line of risk conditions between the bank and the CGI requires human intervention. However, 
the exchange of documents and information between the parties involved is quick and secure, and all the subjects know the progress at any time.

Finally, when the loan contract starts, the bank begins the loan monitoring phase, verifying the regularity of payments and the ongoing creditworthiness of the SME and, if necessary, communicating any changes to the CGIs. As depicted in \cref{fig:enforcement}, the smart contract can be settled to enforce the guarantee if the trigger event occurs, such as when the borrower is unable or unwilling to repay as agreed in the guarantee contract. The smart contract has to predict the eligibility terms of the claim request by the bank according to the predetermined line of risk (the preestablished risk hierarchy). Knowing the risk line in case of default is crucial because if the bank can access the guarantee before other creditors, it may increase moral hazard, reducing the motivation to take action against the borrower. The smart contract can contain predefined conditions and encoded dispute resolution mechanisms. If legal enforcement is necessary, the contract will automatically initiate predefined actions, resulting in efficient and cost-effective resolutions compared to traditional guarantee enforcement. The potential of the blockchain network perfectly matches the need to check the legacy of legal enforcement in this stage more than in the previous onboarding stages. Notably, the fact that different stages of the enforcement process are continuous, transparent, and immutable on the blockchain makes it easier for everyone involved in the credit guarantee process to trust each other. This is because everyone can check the terms and conditions of the agreement and see how it is being carried out in real-time.
Next, we conclude our paper with a summary and discussion of our findings.
\section{Conclusion}
\label{conclusion}
This paper investigates the potential synergies between DLTs and CGSs by proposing a comprehensive framework for applying a blockchain to the credit guarantee life cycle. The framework explores different stages of the process from the CGI's point of view, starting from the initial application to the potential enforcement, when the borrower is unable or unwilling to repay the loan and the bank claims the guarantee. We are unaware of any papers exploring the use of DLTs in the CGS process. We aim to initiate a new research direction by providing detailed information on the procedures, stakeholders, and data security in transactions. Additionally, we examine criteria to select a suitable DLT architecture that can enhance transparency, security, and automation in all stages of CGSs.
We hypothesise to build a private permissionless blockchain that includes the actors involved in the CGS process. 
Considering the characteristics of this kind of blockchain compared to the public ones, only nodes associated with the certified actors can participate, ensuring the privacy of data in exchanging information.
Secondly, we highlight how certain blockchain-based distributed ledger settings match the need to automate, verify, and share KYC data securely and transparently, proceeding without further manual intervention. This application could enhance the KYC process when onboarding a new borrower/guarantee application to the CGS blockchain network, making the process more effective and efficient. 
Moreover, implementing a blockchain network implies ease of use and communication between the borrower and the bank with the CGI. If the smart contract code’s predefined rules and regulations are correctly observed, a slimmer process allows the borrower to easily proceed with the loan request to the bank (also thinking about the possibility of improving the interoperability of the network, including later a smart loan contract). This addresses the double-screening problem that significantly impacts the operational procedures of CGSs by allowing the CGI to establish essential criteria for the borrower's eligibility for the guarantee request. 
Our proposal suggests utilising the blockchain network to notify the guarantee institution when the enforcement trigger event occurs. At this stage, the CGI must determine if the bank's claim request meets the criteria based on the level of risk and legal enforcement history. The network's transparency allows all participants to monitor compliance with terms and conditions, accelerating the process and reducing moral hazard and irregularities.

\bigskip

\noindent\textbf{Acknowledgements.}
This work was partly funded by Sapienza University of Rome under grant RM1221816C0CC0C6 
(\textit{Distributed Ledger and Credit Guarantee Schemes}). L.~Barbaro's work received funding
by the Latium Region under PO~FSE+ grant B83C22004050009 (PPMPP).

\vspace{-2ex}
\bibliographystyle{splncs04}
\bibliography{bibliography}

\end{document}